\documentstyle[12pt]{article}

\newcommand{\noi}{\noindent}
\newcommand{\bc}{\begin{center}}
\newcommand{\ec}{\end{center}}

\def\ifmath#1{\relax\ifmmode #1\else $#1$\fi}
\def\3quarter{{\textstyle{3 \over 4}}}

\def\vs{\vskip}

\def\ra{\rightarrow}

\def\lf{\leaders\hbox to 1em{\hss.\hss}\hfill}

\def\21{$SU(2) \ot U(1)$}
\def\O{\hbox{$\cal O$ }}

\def\etal{\hbox{\it et al., }}
\def\sm{\hbox{standard model }}

\def\neu{\hbox{neutrino }}

\def\neus{\hbox{neutrinos }}

\def\smp{\hbox{standard model. }}

\def\eq#1{{eq. (\ref{#1})}}

\def\VEV#1{\left\langle #1\right\rangle}

\def\lsim{\raise0.3ex\hbox{$\;<$\kern-0.75em\raise-1.1ex\hbox{$\sim\;$}}}
\def\gsim{\raise0.3ex\hbox{$\;>$\kern-0.75em\raise-1.1ex\hbox{$\sim\;$}}}
\def\bel{\begin{letter}}
\def\eel{\end{letter}}
\def\beq{\begin{equation}}
\def\eeq{\end{equation}}
\def\bef{\begin{figure}}
\def\eef{\end{figure}}
\def\bet{\begin{table}}
\def\eet{\end{table}}
\def\bea{\begin{eqnarray}}
\def\ba{\begin{array}}
\def\ea{\end{array}}
\def\bi{\begin{itemize}}
\def\ei{\end{itemize}}
\def\ben{\begin{enumerate}}
\def\een{\end{enumerate}}
\def\ra{\rightarrow}
\def\ot{\otimes}
\def\eea{\end{eqnarray}}

\def\np#1#2#3{           {\it Nucl. Phys. }{\bf #1} (19#2) #3}
\def\pl#1#2#3{           {\it Phys. Lett. }{\bf #1} (19#2) #3}
\def\pr#1#2#3{           {\it Phys. Rev. }{\bf #1} (19#2) #3}
\def\prep#1#2#3{         {\it Phys. Rep. }{\bf #1} (19#2) #3}
\def\prl#1#2#3{          {\it Phys. Rev. Lett. }{\bf #1} (19#2) #3}

\def\n.c.#1#2#3{         {\it Nuovo Cim. }{\bf #1} (19#2) #3}
\def\r.n.c.#1#2#3{       {\it Riv. del Nuovo Cim. }{\bf #1} (19#2) #3}

\relax

\begin{document}
\begin{titlepage}
\pagestyle{empty}
%\rightline{CERN-xxx}
\rightline{FTUV/93-19}
%\rightline{IFIC/93-xxx}
\rightline{April 1993}
\noindent
%\today
%\hfill Submitted to Phys. Lett.\\
\begin{center}
%\hfill prelim. version\\
{\bf MODEL INDEPENDENT HIGGS BOSON MASS LIMITS AT LEP}\\
\vskip 0.4cm
{\bf A. Lopez-Fernandez}
\footnote{Bitnet ALFON@CERNVM}\\
PPE Division, CERN\\
CH-1211 Geneve 23, Switzerland\\
{\bf J. C. Rom\~ao}
\footnote{Bitnet ROMAO@PTIFM}\\
{Departamento de F\'{\i}sica, Instituto Superior T\'ecnico\\
Av. Rovisco Pais, 1 - 1096 Lisboa Codex, PORTUGAL}\\
and \\
{\bf F. de Campos}
\footnote{Bitnet CAMPOSC@vm.ci.uv.es - Decnet 16444::CAMPOSC}
{\bf and J. W. F. Valle}
\footnote{Bitnet VALLE@vm.ci.uv.es - Decnet 16444::VALLE}\\
Instituto de Fisica Corpuscular - IFIC/CSIC\\
Dept. de F\'isica Te\`orica, Universitat de Val\`encia\\
46100 Burjassot, Val\`encia, SPAIN\\
\vskip 0.4cm
{\bf ABSTRACT}\\
\end{center}
\vskip 0.1cm
\noi
We derive model-independent constraints on Higgs
mass and couplings from the present LEP1 data
samples and discuss the prospects for detecting
the associated signals for higher masses,
accessible at LEP2. This work is motivated
by the fact that, in many extensions of the
standard model, the Higgs boson can have
substantial "invisible" decay modes, for
example, into light or massless weakly
interacting Goldstone bosons associated
to the spontaneous violation
of lepton number below the weak scale.

\vs 1cm
%\begin{flushleft}
%CERN-TH.6652/92\\
%September 1992
%\end{flushleft}
\vfill
\noi
\end{titlepage}
\setcounter{page}{1}
\pagestyle{plain}

\section{Introduction}
%\hspace{\parindent}

The problem of mass generation remains one
of the main puzzles in particle physics today.
In the standard model all masses arise as a result
of the spontaneous breaking of \21 the gauge symmetry.
This implies the existence of an elementary Higgs boson
\cite{HIGGS}, not yet found. Recently the LEP
experiments on  $e^+ e^-$ collisions around the
Z peak have placed important restrictions on
the Higgs boson mass
\beq
\label{1}
m_{H_{SM}} \gsim 60 \rm{GeV}.
\eeq
This limit holds in the \smp

There are many reasons
to think that there may exist additional Higgs bosons
in nature. One such extension of the minimal \sm is
provided by supersymmetry and the desire to tackle the
hierarchy problem \cite{revsusy}. There are, however,
many other motivations. One is the question of \neu
masses, whose existence is
presently suggested by astrophysical
data on solar and atmospheric \neus as well as
cosmological observations related to the large
scale structure of the universe and the possible
need for hot dark matter \cite{dallas}. Most
extensions of the minimal standard model
to induce \neu masses require an enlargement in the
Higgs sector \cite{fae}. Another motivation to
extend the Higgs sector is to generate the
observed baryon excess by electroweak physics
\cite{Kuz}. Indeed, the latter requires
$m_{H_{SM}} \lsim 40$ GeV \cite{Dineetal92a}
in conflict with \eq{1}. This limit can be
avoided in models with new Higgs bosons
\cite{BGLAST,2Higgs}.

Amongst the extensions of the \sm which have been
suggested to generate \neu masses, the majoron
models are particularly interesting and have been
widely discussed \cite{fae}. The majoron is a Goldstone
boson associated with the spontaneous breaking of the lepton
number. Astrophysical arguments based from stellar cooling rates
constrain its couplings to the charged fermions
\cite{KIM}, while the LEP measurements of the invisible Z width
restrict the majoron couplings to the gauge bosons in an important
way. In particular, models where the majoron is not a
singlet \cite{GR} under the \21 symmetry are now
excluded \cite{LEP1}.

There is, however, a wide class of models
\cite{JoshipuraValle92}, motivated by \neu physics,
which are characterized  by the spontaneous violation
of a global $U(1)$ lepton number symmetry by an \21
singlet vacuum expectation value $\VEV{\sigma}$ \cite{CMP}.
These models may naturally explain the \neu masses required
by astrophysical and cosmological observations.
Another example is provided by supersymmetric
extensions of the \sm where R parity is
spontaneously violated \cite{MASI}.

In all these extensions of the minimal standard model
the global $U(1)$ lepton number symmetry is spontaneously
violated close to the electroweak scale.
Such a low scale for the lepton number violation
is preferred since, in these models, $m_\nu \to 0$
as $\VEV{\sigma} \ra 0$.
As a result, a relatively low value of $\VEV{\sigma}$ is
required in order to obtain naturally small neutrino
masses. These may arise either at the tree level
or radiatively \cite{JoshipuraValle92}.

An alternative argument for why the violation of
a global symmetry should happen at a relatively
low scale has recently been given. It states that,
in the presence
of nonperturbative gravitational effects, global
symmetries are generally broken explicitly, so
that any related Goldstone boson, such as the majoron,
is expected to acquire a small mass by gravitational
effects. While the corresponding majoron mass is lower
than a keV or so, it could affect the evolution of the
universe. As a result, the majoron must be unstable, to avoid
conflict with cosmology. This leads to an upper
bound on the lepton breaking scale
$\VEV{\sigma} \lsim \O(10)$ TeV \cite{Goran92}.

In any model with a spontaneous violation of a
global $U(1)$ symmetry around the weak scale
(or below) the corresponding Goldstone boson
has significant couplings to the Higgs bosons,
even if its other couplings are suppressed.
This implies that the Higgs boson can decay
with a substantial branching ratio into the
invisible mode \cite{JoshipuraValle92,Joshi92,HJJ}
\beq
h \ra J\;+\;J
\label{JJ}
\eeq
where $J$ denotes the majoron.

Such an invisible Higgs decay would lead to events
with large missing energy that could be observable
at LEP and affect the corresponding Higgs mass bounds.

It is the purpose of this letter to
derive {\sl in a model independent way}
the limits on the Higgs boson mass that
can be deduced from the present LEP samples.
For simplicity we focus on the simplest model,
sketched in section 2. We obtain limits that
must hold {\sl irrespective of whether
the Higgs decays visibly or invisibly}. In order
to do this we first determine the lightest Higgs
boson production rates. These are, generically,
somewhat suppressed with respect to the \sm
prediction. We call this suppression factor
$\epsilon^2$. Then we combine three final-state
search  methods:
\ben
\item
$Z \ra H Z^*$, $H \ra q\bar{q}$, $Z \ra \nu \nu$ or $ll$
where we directly use the SM Higgs search results
\item
$Z \ra H Z^*$, $H \ra $invisible, $Z \ra l l$,
where we combine the results of acoplanar
lepton pair searches. This gives good limits
for low Higgs masses
\item
$Z \ra  H Z^*$, $H \ra$ invisible, $Z \ra q\bar{q}$
where we reinterpret the results SM Higgs search in
the $H \nu \nu$ channel. This allows better limits
for high values of the Higgs mass.
\een
Our results are summarized in figures 1 and 2.
Finally, we have also determined the additional
range of parameters that can be covered by LEPII
for a total integrated luminosity of 500 pb$^{-1}$
and centre-of-mass energies of 175 GeV and 190 GeV.

\section{The Simplest Example}

There are many models of interest for
\neu physics, astrophysics and cosmology
where the Higgs boson will have important
invisible decay rates. Some examples have
been considered previously \cite{JoshipuraValle92,Joshi92}.
For our present purposes they do not need to be
specified beyond the structure of their neutral
scalar potential responsible for the breaking
of the \21 and the global symmetries.

The simplest model contains, in addition to
the scalar Higgs doublet of the \sm an
additional complex singlet $\sigma$
which also acquires a nonzero
vacuum expectation value $\VEV{\sigma}$
which breaks the global symmetry.
The scalar potential is given by
\cite{JoshipuraValle92,Joshi92}
\begin{eqnarray}
\label{V1}
V = \mu_{\phi}^2\phi^{\dagger}\phi
+\mu_{\sigma}^2\sigma^{\dagger}\sigma
+ \lambda_{1}(\phi^{\dagger}\phi)^2+
 \lambda_2
(\sigma^{\dagger}\sigma)^2
+\delta (\phi^{\dagger}\phi)(\sigma^{\dagger}\sigma)
\end{eqnarray}
Terms like $\sigma^2$ are omitted above in view of the imposed
$U(1)$ invariance under which we require $\sigma$ to transform
nontrivially and $\phi$ to be trivial. Let $\sigma \equiv
\frac{w}{\sqrt 2}+\frac{R_2+iI_2}{\sqrt{2}}$, $\phi^0\equiv
\frac{v}{\sqrt 2}+\frac{R_1+iI_1}{\sqrt {2}}$, where we have set
$\VEV{\sigma} =\frac{w}{\sqrt{2}}$ and $\VEV{\phi^0}=\frac{v}{\sqrt{2}}$.
The above potential then leads to a physical massless Goldstone
boson, namely the majoron $J \equiv {\rm Im}\; \sigma$ and two
massive neutral scalars $H_i$ ($i$= 1,2)
\beq
H_i={\hat O}_{ij}\;R_j
\eeq
The mixing ${\hat O}$ can be parametrized as
\begin{equation}
{\hat O}=\left( \begin{array}{cc}
cos\theta&sin\theta\\
-sin\theta&cos\theta\\
\end{array} \right)
\end{equation}
mixing angle $\theta$ as well as the Higgs masses $M_i^2$
are related to the parameters of the potential in the
following way:
\begin{eqnarray}
\label{teta}
2\delta v w &=& (M^2_2-M^2_1) \sin 2 \theta \nonumber\\
2 \lambda _1 v^2&=&M^2_1 \cos ^2\theta+M^2_2 \sin ^2 \theta \nonumber\\
2 \lambda _2 w^2&=&M^2_2 \cos ^2\theta+M^2_1 \sin ^2 \theta. \nonumber\\
tan2\theta&=&-\frac{\delta v \omega}{\lambda_1v^2 - \lambda_2\omega^2}
\end{eqnarray}
We can take the physical masses $M_{1,2}^2$,
the mixing angle $\theta$, and the ratio
of two vacuum expectation values
characterizing the violation of the \21
and global symmetries,
\beq
\tan  \beta =\frac{v}{w}
\eeq
as our four independent parameters. In terms of
these all the relevant couplings, Higgs boson
production cross sections and decay rates
can be fixed.

This completes the discussion on the Higgs boson
spectrum and couplings in this simplest scheme.
Note that there are no physical charged Higgs
bosons in this case. In more complicated models,
e.g. supersymmetric ones \cite{MASI}, there may
exist also massive CP-odd scalar bosons, as well
as electrically charged bosons. For simplicity
we will not consider this case in what follows.

\section{Higgs Production and Decay}

The Higgs boson can be produced at the $e^+ e^-$ collider
through its couplings to $Z$. In the simplest prototype
model sketched above only the doublet Higgs boson $\phi$ has a
coupling to the Z in the weak basis, not the \21 singlet
field $\sigma$. After diagonalizing the scalar boson mass
matrix one finds that the two CP even mass eigenstates $H_i$
($i$=1,2) have couplings to the Z, involving the mixing
angle $\theta$. These couplings may be given as follows
\begin{equation}
\label{HZZ1}
{\cal L}_{HZZ}
=(\sqrt 2 G_F)^{1/2} M_Z^2 Z_{\mu}Z^{\mu}{\hat O}_{i1}H_i
\end{equation}
Through these couplings both CP even Higgs bosons may
be produced through the Bjorken process. As long as the
mixing appearing in \eq{HZZ1} is $\O$(1), all Higgs
bosons can have significant production rates, but
always smaller than in the standard model. For
example, if only the light field $H_1$ is below
the Z boson mass, only this one will be produced,
with a rate $cos^2\theta$ smaller than in the
\sm.

We now turn to the Higgs boson decay rates,
which are sensitive to the details of the
mass spectrum and Higgs potential.  For
definiteness we focus on the simplest
potential, given in \eq{V1}. In this case
the coupling of $H_i$ to the majoron $J$
can be written in the following way:
\begin{equation}
\label{J1}
{\cal L}_J=
\frac{(\sqrt 2 G_F)^{1/2}}{2}\tan  \beta[M_2^2 cos \theta H_2-M_1^2 sin
\theta H_1]J^2
\end{equation}
The width for the invisible $H_i$ decay  can be
parametrized by
\begin{equation}
\label{HJJ}
\Gamma(H\rightarrow JJ)=\frac{\sqrt 2 G_F}{32 \pi} M_{H_i}^3 g^2_{H_i JJ}
\end{equation}
where the corresponding couplings are given by
\begin{equation}
\label{HJJ1}
g_{H_iJJ}=tan\beta\;\;\;{\hat O}_{i2}
\end{equation}
The rate for $H\rightarrow b \overline b$ also gets diluted compared
to the standard model prediction, because of the mixing effects.
Explicitly one has,
\begin{equation}
\Gamma(H\rightarrow b \overline b)=\frac{3\sqrt 2 G_F}{8
\pi}M_Hm_b^2(1-4m_b^2/M_H^2)^{3/2}g^2_{Hb\overline b}
\end{equation}
which is smaller than th \sm prediction by the factor
$g_{H_ib\overline b}$, where
\begin{equation}
\begin{array}{ccc}
g_{H_ib\overline b}= {\hat O}_{i1}\;\;\;\;
\end{array}
\end{equation}

The  width of the Higgs decay to the $JJ$ relative to
the conventional $b\overline b$ mode depends upon the
mixing angles. There are, in principle, three cases
to consider: ($i$) $\omega\approx v$, ($ii$) $\omega\gg v$
and ($iii$) $\omega \ll v$. In the first case, one can
see from \eq{teta} that the mixing among the doublet and
singlet field can be substantial if the parameters of the
quartic terms in the Higgs potential are of comparable
magnitude.  As a result, the production as well as the
invisible decay of both physical Higgs bosons $H_i$
can be important. In this case we have
\begin{eqnarray}
\label{ratio}
\frac{\Gamma(H_1 \rightarrow JJ)}{\Gamma(H_1\rightarrow b\overline b)}&=&
\frac{1}{12}\left(\frac{M_1}{m_b}\right)^2
(1-4m_b^2/M_1^2)^{-3/2}(tan\beta\:tan\theta)^2 \nonumber\\
&\approx& 8\left(\frac{M_1}{50 GeV}\right)^2
(tan\beta\:tan\theta)^2
\end{eqnarray}
A similar expression with $\tan\theta$ replaced by $\cot\theta$ holds
in case of $H_2$. It is clear that a Higgs boson with $M_H>50$ GeV
decays mostly invisibly if $\tan\beta$ and $\tan\theta$ are $\O$(1).
The production of $H_1(H_2)$ gets diluted compared to the standard
model prediction by $cos^2\theta$ ($\sin^2\theta)$.

If $\omega$ and $v$ are very different from each other then
the mixing angle in \eq{teta} is very small. Hence in cases
$(ii)$ and $(iii)$, only the predominantly doublet component
($H_1)$ will be produced. Using \eq{teta} in the basic majoron
coupling, \eq{J1}, one finds that
if $\omega\gg v$ then mostly the singlet
Higgs boson which decays to two majorons.
But its production rate is, of course, negligible.
In contrast, for the other case, $\omega \ll v$,
the doublet Higgs field mainly decays to majorons
and is produced without any substantial suppression
relative to the standard model predictions.

In summary, the invisible Higgs decay mode is expected
to have quite important implications if there exists,
as suggested by \neu physics, a global symmetry that
gets broken around or below the weak scale, not too
much above. From this point of view it is therefore
desirable to obtain limits on Higgs bosons that are
not vitiated by detailed assumptions on its mode
of decay.

\section{LEP I Limits}

The production and subsequent decay of any Higgs boson which
may decay visibly or invisibly involves three independent
paramenters: the Higgs boson mass $M_H$, its coupling
strength to the Z, normalized by that of the \sm, we call
this factor $\epsilon^2$, and the invisible Higgs boson
decay branching ratio.

We have used the results published by the LEP experiments on the
searches for various exotic channels in order to deduce the regions
in the parameter space of the model that can be ruled out already.
The procedure was the following.
For each value of the Higgs mass, we calculated the
lower bound on $\epsilon^2$, as a function of the
branching ratio $BR(H \ra $ visible). By taking
the highest such bound for $BR(H \ra $ visible)
in the range between 0 and 1, we obtained the absolute bound
on $\epsilon^2$ as a function of $M_H$.

For a Higgs of low mass (below 30 GeV) decaying to invisible
particles we considered the process $Z \ra H Z^*$,
with $Z^* \ra e^+e^-$   or $Z^* \ra \mu^+\mu^-$
and combined the results of the LEP experiments on
the search for acoplanar lepton pairs \cite{OPAL91,ALEPH91,L3_91}
which found no candidates in a total sample corresponding to
780.000 hadronic Z decays. The efficiencies for the detection
of the signal range from 20\% at very low Higgs masses to
almost 50\% for $M_H = 25$ GeV.

For higher Higgs masses the rate of the process used above is too
small, and we considered instead the channel $Z  \ra HZ^*$,
$Z^*  \ra q\bar{q}$.
Here we translated the results of the searches for the
Standard Model Higgs in the channel $Z  \ra Z^* H_{SM}$,
with $H_{SM} \ra q\bar{q}$ and $Z \ra  \nu \bar{\nu}$,
following ref. \cite{Felcini92}. The efficiency of these
searches for an invisible Higgs increases from 25\% at
$M_H = 30$ GeV to about 50\% at $M_H = 50$ GeV.

For visible decays of the Higgs boson its signature is the same
as that of the Standard Model one, and the searches for this particle
can be applied directly. For masses below 12 GeV we have taken the results
of a model independent analysis made by the L3 collaboration
(ref. \cite{L3_92}).
For masses between 12 and 35 GeV we combined the results from
references  \cite{OPAL91,Felcini92,L3_92}; finally for
masses up to 60 GeV we used the combined result of all the
four LEP experiments given in reference \cite{Felcini92}.
In all cases the bound on the ratio $BR(Z \ra ZH)/BR(Z \ra ZH_{SM})$
was calculated from the quoted sensitivity, taking into account
the background events where they existed.%%(MORE DETAIL?)

As an illustration we show in Figure 1 the exclusion contours
in the plane $\epsilon^2$
vs. $BR(H \ra $ visible) for the particular choice for the Higgs
mass   $M_H = 50$ GeV. The two curves corresponding to the searches
for visible and invisible decays are combined to give the final bound;
values of $\epsilon^2$ above 0.2 are ruled out independently of the
value of $BR(H \ra $ visible). The solid line in Figure 2 shows
the region in the $\epsilon^2$ vs. $M_H$ that can be excluded by
the present LEP analyses, independent of the mode of Higgs decay,
visible or invisible.

\section{Prospects for LEPII}

We have also estimated the additional range of parameters that can
be covered by LEPII. We assumed that the total luminosity collected
will be 500 pb$^{-1}$, and give the results for two values of the
centre-of-mass energy: 175 GeV and 190 GeV.

Our results on the visible decays of the Higgs are based on the
study of efficiencies and backgrounds in the search for the Standard
Model Higgs described in reference \cite{Janot92}.  For the invisible
decays of the Higgs we considered only the channel  HZ with
$Z \ra e^+e^-$   or $Z \ra \mu^+\mu^-$, giving a signature of two leptons
plus missing transverse momentum. The requirement that the invariant
mass of the two leptons must be close to the Z mass can kill most of
the background from WW and $\gamma\gamma$ events; the background from
ZZ events with one of the Z decaying to neutrinos is small and the
measurement of the mass recoiling against the two leptons allows
to further reduce it, at least for $M_H$ not too close to $M_Z$.
Hadronic decays of the Z were not considered, since the background
from WW and $We\nu$ events is very large, and b-tagging is much less
useful than in the search for $Z H_{SM}$ with
$Z \ra  \nu \bar{\nu}$, since the $Zb\bar{b}$ branching ratio
is much smaller than $Hb\bar{b}$ in the \smp

The dashed and dotted curves on figure 2 show
the exclusion contours in the $\epsilon^2$ vs. $M_H$ plane
that can be explored at LEPII, for the given centre-of-mass energies.
Again, these contours are valid irrespective of whether
the Higgs decays visibly, as in the \sm, or invisibly.

\section{Discussion}

The Higgs can decay to a pair of invisible massless
Goldstone bosons in a wide class of models in which
a global symmetry, such as lepton number, is broken
spontaneously around or below the weak scale.
These models are attractive from the point
of view of \neu physics and suggest the need
to search for the Higgs boson in the invisible
mode.

We have presented {\sl model independent limits}
on the Higgs boson mass and HZZ coupling strength
that can be deduced from the present LEP samples.
Our limits combine three final-state searches
and are summarized in figures 1 and 2.
These limits do not depend on the mode
of Higgs boson decay. They are probably
conservative and could still be somewhat
improved with more data and/or more refined
analysis. They apply to a very wide class of
extensions of the \sm, including many models
where \neus acquire mass as a result of the
spontaneous violation of lepton number around
or below the weak scale.  Other global symmetries,
such as the Peccei-Quinn symmetry, are not
relevant in this context. Moreover, we
mention that there are other ways
to engender invisible Higgs decays,
e.g. in the minimal supersymmetric \sm,
in which the Higgs boson may decay as
$H \ra \chi \chi$ where $\chi$ is the
lightest neutralino. This would require
$2 m_\chi < M_H$.

Apart from the invisible Higgs boson decay,
the possible validity of these majoron schemes
may have important consequences for the physics
of \neus and weak interactions \cite{fae,granada}.

The possibility of invisible Higgs decay
is also very interesting from the point of
view of a linear $e^+ e^-$ collider at higher
energy \cite{EE500}. Heavier, intermediate-mass,
Higgs bosons can also be searched at high energy
hadron supercolliders such as LHC/SSC \cite{bj,kane}.
The limits from LEP derived in this paper should
serve as useful guidance for such future searches.

%%\vfill
This work was partially supported by
Accion Integrada Hispano-Portuguesa under
grant N. HP-48B. The work of F.
de Campos was supported by CNPq (Brazil).

\newpage

\section*{Figure Captions}
\noindent
{\bf Figure 1}\\
Figure 1 gives shows the exclusion contours in the plane $\epsilon^2$
vs. $BR(H \ra $ visible) for the particular choice $m_H = 50$ GeV.
The two curves corresponding to the searches for visible (curve A)
and invisible (curve B) decays are combined to give the final
bound, which holds irrespective of the value of $BR(H \ra $ visible).\\
\noindent
{\bf Figure 2}\\
The solid curve shows the region in the $\epsilon^2$ vs. $m_H$ that can be
excluded by the present LEP analyses, independent of the mode of Higgs decay,
visible or invisible. The dashed and dotted curves on figure 2 show
the exclusion contours in the $\epsilon^2$ vs $m_H$ plane that
can be explored at LEPII, for the given centre-of-mass energies.

\end{document}